\documentclass[aps,a4paper,floatfix,nofootinbib]{revtex4}

\usepackage[utf8]{inputenc}

\usepackage{graphicx,color}
\usepackage{amsmath,amssymb,amsfonts,amsthm,amscd,bm}
\usepackage{booktabs}
\usepackage{cancel}

\def\tilde{\widetilde}
\def\bar{\overline}
\def\hat{\widehat}
\def\*{\star}
\def\[{\left[}
\def\]{\right]}
\def\({\left(}      

\def\){\right)}

\def\frac#1#2{\dfrac{#1}{#2}}
\def\inv#1{\dfrac{1}{#1}}
\def\half{\tfrac{1}{2}}
\def\d{\partial}

\def\2pi{\hbox{$2\pi i$}}

\def\dsl{\raise.15ex\hbox{/}\kern-.57em\partial}
\def\Dsl{\,\raise.15ex\hbox{/}\mkern-.13.5mu D}
\def\CA{{\cal A}}      
      \def\CF{{\cal F}}
\def\CG{{\cal G}}      
      
      \def\CO{{\cal O}}
      
\def\CS{{\cal S}}   \def\CT{{\cal T}}   \def\CU{{\cal U}}

\def\2pi{\hbox{$2\pi i$}}

\def\dsl{\raise.15ex\hbox{/}\kern-.57em\partial}
\def\Dsl{\,\raise.15ex\hbox{/}\mkern-.13.5mu D}
\font\numbers=cmss12
\font\upright=cmu10 scaled\magstep1
\def\stroke{\vrule height8pt width0.4pt depth-0.1pt}
\def\topfleck{\vrule height8pt width0.5pt depth-5.9pt}
\def\botfleck{\vrule height2pt width0.5pt depth0.1pt}
\def\Zmath{\vcenter{\hbox{\numbers\rlap{\rlap{Z}\kern
    0.8pt\topfleck}\kern 2.2pt
    \rlap Z\kern 6pt\botfleck\kern 1pt}}}
\def\Qmath{
    \vcenter{\hbox{\upright\rlap{\rlap{Q}\kern3.8pt\stroke}\phantom{Q}}}}
\def\Nmath{\vcenter{\hbox{\upright\rlap{I}\kern 1.7pt N}}}
\def\Cmath{\vcenter{\hbox{\upright\rlap{\rlap{C}\kern
                   3.8pt\stroke}\phantom{C}}}}
\def\Rmath{\vcenter{\hbox{\upright\rlap{I}\kern 1.7pt R}}}
\def\Z{\ifmmode\Zmath\else$\Zmath$\fi}
\def\Q{\ifmmode\Qmath\else$\Qmath$\fi}
\def\N{\ifmmode\Nmath\else$\Nmath$\fi}
\def\C{\ifmmode\Cmath\else$\Cmath$\fi}
\def\R{\ifmmode\Rmath\else$\Rmath$\fi}
\def\barray{\begin{eqnarray}}
\def\earray{\end{eqnarray}}
\def\beq{\begin{equation}}
\def\eeq{\end{equation}}

\def\kvec{{\bf{k}}}

\def\Li{{\rm Li}}

\def\AA{\leavevmode\setbox0=\hbox{h}
\dimen0=\ht0 \advance\dimen0 by-1ex\rlap{\raise.67\dimen0\hbox{\char'27}}A}

\def\Li{{\rm Li}}

\makeatletter
\def\iddots{\mathinner{\mkern1mu\raise\p@
\vbox{\kern7\p@\hbox{.}}\mkern2mu
\raise4\p@\hbox{.}\mkern2mu\raise7\p@\hbox{.}\mkern1mu}}
\makeatother

\def\Li{{\rm Li}}

\theoremstyle{plain}

\theoremstyle{remark}

\newtheorem{assumption}{Assumption}

\def\alphavec{\bm{\alpha}}
\def\phivec{\bm{\phi}}

\def\gcal{\mathfrak{g}}
\def\pcal{\mathfrak{p}}

\def\thetatilde{\tilde{\theta}}

\def\rhovac{{\rho_{\rm vac}}}

\def\Lvac{\langle 0 |}
\def\Rvac{| 0 \rangle}

\def\Z{\mathbb{Z}}

\def\kvec{{\bf k}}
\def\xvec{{\bf x}}

\def\Fmin{F_{\rm min}}

\def\TraceT{\Theta}

\def\mhat{\hat{m}}

\def\mgod{m_{\rm z}}

\def\zeron{{\rm zeron}}

\def\Dsmall{{\scriptscriptstyle{D}}}

\begin{document}

\title{Vacuum energy density from the form factor bootstrap.}

\author{
 Andr\'e  LeClair\footnote{andre.leclair@cornell.edu} 
}
\affiliation{Cornell University, Physics Department, Ithaca, NY 14850,  USA} 

\begin{abstract}

The form-factor bootstrap is incomplete until one normalizes the zero-particle form factor. 
For the stress energy tensor we describe how to obtain the vacuum energy density $\rhovac$,    defined as 
$\langle 0| T_{\mu\nu} | 0 \rangle  = \rho_{\rm vac}  \, g_{\mu\nu}$,    from the form-factor bootstrap.  
Even for integrable QFT's in D=2 spacetime dimensions,  this prescription is new,  although it reproduces previously known results obtained
in a different and more difficult thermodynamic Bethe ansatz computation.       We propose a version of this prescription in D=4 dimensions. 
For these even dimensions,   the vacuum energy density has the universal form $\rho_{\rm vac}  \propto m^D/\mathfrak{g}$ where $\mathfrak{g}$ is a dimensionless interaction coupling constant which can be determined  from the high energy behavior of the S-matrix.    In the limit $\mathfrak{g} \to 0$,   $\rho_{\rm vac} $ diverges due to well understood  UV divergences in free quantum field theories.   
If we assume the the observed Cosmological Constant originates from the vacuum energy density  $\rho_{\rm vac}$  computed as proposed here,  then this suggests there  must exist a particle which does not obtain its mass from spontaneous symmetry breaking in the electro-weak sector,   which we designate as  the ``zeron".    A strong candidate for the zeron is a massive Majorana neutrino.

\end{abstract}

\maketitle
\tableofcontents

\section{Introduction}

 Bootstrap ideas,  which originated in the 1960's as an attempt to understand the strong interactions,    
 have found great success in D=2  spacetime dimensions,  in particular for  conformal field theory (CFT) \cite{BPZ}  and integrable massive theories \cite{ZamoZamo0}.    
In recent years the bootstrap has been developed in some detail in higher dimension D with surprising success.    For CFT's see the 
review  \cite{Poland}.      More recently the bootstrap has been studied for massive theories \cite{Paulos1,Homrich,Karateev} and the latter is more relevant to the present work.

In this article we are primarily concerned with the vacuum energy density $\rhovac$ defined as a vacuum expectation value of the 
stress-energy tensor $T_{\mu\nu}$:
\beq
\label{rhvacdef0}
 \Lvac T_{\mu\nu} \Rvac  =  \rhovac \, g_{\mu\nu},
\eeq 
with the convention $\{ g_{\mu \nu} \} = {\rm diag}  \{ 1, -1, -1, -1, \ldots  \}$.   
From this definition of the vacuum energy density $\rhovac$  one  has 
\beq
\label{rhovacTrace}
\rhovac =  \Lvac \TraceT \Rvac  /D, ~~~~~ \TraceT = T_\mu^\mu.
\eeq
We thus focus on form factors for the trace of the stress energy tensor $\TraceT$.     For CFT's,   $\TraceT = 0$.    
The form factor bootstrap in principle relates $n$-particle form factors to those with $n-2$ particles,   thus we address the problem of
determining $\rhovac$ from the 2-particle form factor.      For integrable theories in $D=2$ the form-factor bootstrap is very well developed 
\cite{Smirnov} with many applications \cite{MussardoBook}.   In any dimension,   even in $2D$,   the form factor bootstrap  for any operator 
$\CO$ is incomplete since
the basic equations are linear in the form factor and are invariant under a rescaling by an arbitrary constant.  Thus the form factors for any operator $\CO$ are incomplete until one specifies its vacuum expectation value $\Lvac \CO \Rvac$.       This necessarily comes from additional ultra-violet (UV) data such as the thermodynamic Bethe ansatz (TBA).      A main result of this paper is a prescription for determining $\Lvac \TraceT \Rvac$ from the 2-particle form factor.      This prescription is new,  even in $2D$,   and as we will show it reproduces known results for $\rhovac$ previously obtained from the TBA,  but without introducing a finite temperature.    More importantly it has a natural generalization to D=4.

In \cite{ALCC} a non-perturbative definition of $\rhovac$ was proposed based on thermodynamics.    In the ``Thermal" channel euclidean time is compactified on a circle of circumference $\beta = 1/T$ where $T$ is the temperature.    Thus states are the usual scattering states in infinite 
$d=D-1$ dimensional volume.    This is in contrast to the spatially compactified channel ``SpC",  where states  partially live on a circle of one compactified spatial dimension.     We showed that these two channels lead to the same result for free massive theories,  however the Thermal channel is better behaved since most integrals are already convergent.     The results of this paper are implicitly based on the Thermal channel in the limit $\beta \to 0$.    Let us summarize the non-perturbative definition of $\rhovac$ in the thermal channel proposed in \cite{Mingling,ALCC}. 
Let $p$ denote the pressure and $\CF$ the free energy density.    Assuming the theory has a single mass scale $m$,  such as the mass of the lightest particle,  we define the scaling variable
$r=m\beta$.    Then one can write 
\beq
\label{cDdef}
p\,  \beta^\Dsmall = - \beta^\Dsmall \CF \equiv \chi (D) \, c(r), ~~~~ \chi(D)  \equiv \pi^{-\Dsmall /2} \Gamma(D/2) \zeta (D).
\eeq
The function $\chi (D)$ is the famous combination discovered by Riemann that satisfies the functional equation $\chi(D) = \chi(1-D)$ which is essential for
establishing Modularity for the Thermal  verses SpC channels \cite{ALCC}.   
If the theory is UV complete and a CFT  then one expects
\beq
\label{cofr}
c(r) = c_{\rm uv} + c_\Dsmall  \, r^\Dsmall + \ldots 
\eeq
where $\ldots = \sum_\pcal c_\pcal \,  r^\pcal $ refers to terms with different powers of $r$ coming from perturbation theory  about the CFT by 
 relevant operators. 
 Then we proposed
\beq
\label{rhovacvac}
\rhovac = - c_\Dsmall  \chi (D) m^\Dsmall .
\eeq
The normalization of $c(r)$ is such that $c_{\rm uv} = 1$ for a free massless boson for any $D$.   In $D=4$ dimensions one expects the various powers $\pcal$ to be irrational and shouldn't mix with $c_\Dsmall$, unlike in 2D where conformal perturbation theory can lead to divergent terms with
powers $r^2$ due to the rationality of anomalous dimensions of the relevant perturbation about the UV CFT.    
In the Thermal channel,    the $c_\Dsmall$ term is difficult to calculate even with the TBA,    and this provides the main motivation for this article,
namely to determine $\rhovac$ without having to calculate the full free energy density $\CF$ at arbitrary temperature $T$.    Our prescription below can thus be viewed as extracting $\rhovac$ from a  proper $\beta \to 0$,  i.e. high temperature  (UV) limit. 
For integrable quantum field theories (QFT)'s with diagonal scattering,  using the TBA one can find a simple formula for $\rhovac$ originally due to Al. Zamolodchikov 
\cite{ZamoTBA,KlassenMelzer}
\beq
\label{rhovacZamo}
\rhovac =  \frac{m_1^2}{2 \gcal}, 
\eeq
where $m_1$ is the lightest particle and $\gcal$ is a finite dimensionless interaction coupling which can be extracted from the S-matrix for the scattering of 
$m_1$ with itself (see below for a review).        
Using  D=4 Lattice QCD results for two light quarks and one massive one (the strange quark)  we used \eqref{rhovacvac} to estimate $\rhovac$ and obtained the reasonable value $\rhovac \approx -(200 {\rm MeV} )^4$,  and this calculation relied on asymptotic freedom \cite{Wilczek,Gross}.

A main goal of this article is to obtain a formula analogous to \eqref{rhovacZamo} in higher dimensions directly from the S-matrix and the form factor bootstrap.   Towards this the following remark is important.      For free particles there is an unavoidable divergence in the UV limit $\beta \to 0$ for D even.     Specifically,  for free theories ($\gcal = 0$) one finds \cite{ALCC} 
\beq
\label{rhoFree}
\rhovac =  \pm 
\begin{cases}
\frac{m^2}{4 \pi} \(  \log (4\pi/m\beta )  + \half - \gamma_E  \) ~~~~~~~(D=2) \\ 
\\
\frac{m^4}{32 \pi^2} \( \log(4\pi/m\beta)   + \tfrac34 -\gamma_E \) ~~~~~(D=4)
\end{cases}
\eeq
where $+/-$ corresponds to bosons/fermions and $\gamma_E$ is the Euler constant.    Remarkably the result \eqref{rhovacZamo} shows that interactions,  i.e.  $\gcal \neq 0$, 
can regularize the divergence as $\beta \to 0$ in the free theory.     This implies that {\it UV divergences in the non-interacting theory can be absorbed into physical masses $m_{\rm phys}$ in order obtain a physical $\rhovac$.}    The zero temperature S-matrix depends only on these physical masses and coupling $\gcal$ and should render a finite result for $\rhovac$.       A main result of this article is such a formula for $D=4$,  and based on this discussion we expect that
$\lim_{\gcal \to 0}  \rhovac = \infty$.    Henceforth $m$ refers to these physical masses.

One aspect of our analysis is the principle of a  {\it particular democracy}.    Namely since the vacuum $\Rvac$ has no quantum numbers and cannot be excited as a resonance in scattering,   then in principle the scattering of  {\it any} particle  with it's anti-particle can probe $\rhovac$,   and they should all agree\footnote{``Every particle is created equal when it comes to experiencing the void."}.    For $D=2$ integrable theories we will 
show how this principle follows from the S-matrix bootstrap in the specific example of Toda theories based on the affine extension of the the Lie group
$SU(N+1)$.     

We present our results as follows.   In the next section we describe very general properties of the 2-particle form factors of the stress-energy tensor which apply to any dimension D and don't rely on the bootstrap nor integrability.
    This  section mainly  just serves to fix  the normalization of the form factors in terms of {\it physical} masses.        
In Section III we turn to D=2 integrable theories and propose a prescription for how to obtain $\rhovac$ from the 2-particle form factor based on the bootstrap.   Although the final result for $\rhovac$ is known to be of the form \eqref{rhovacZamo} by other methods in particular the TBA,   our derivation directly from the S-matrix without relying on the TBA is new and actually simpler.    We present the example of the affine Toda theories
for SU(N+1)  in order to explain how the {\it particular democracy}  originates from the 
S-matrix bootstrap.    The $SU(2)$ case is the sinh-Gordon model and we carry out a detailed analysis in this case using the explicit formula for the 
2-particle form-factor.    
In Section IV we take some steps towards higher dimensions.        There we first rewrite  the $D=2$ result using the Mandelstam variable $s$.
This leads us to propose a formula for $\rhovac$ in $D=4$ based entirely on the high energy behavior of the S-matrix.     
Since $\rhovac$ is central to any discussion of the Cosmological Constant Problem,   in Section V we speculate on  potential implications for
physics beyond the Standard Model.    We argue that if $\rhovac$ is enough to explain the observed value of the cosmological constant 
$\rho_\Lambda$,   then this requires at least one stable massive particle in the deep UV above spontaneous symmetry breaking of the electro-weak sector.   With lack of a better name we refer  to such a hypothetical particle as the ``zero-on" , where ``zero" refers to the vacuum 
$\Rvac$,   with mass $\mgod$.    In order to explain the measured value of $\rho_\Lambda$,  it turns out  that a massive Majorana neutrino is a good candidate for the \zeron,   since it is thought that its mass cannot yet be explained by the Higgs mechanism,  and its  proposed mass based on phenomenology is in the right ball-park to explain the observed value of the Cosmological Constant.

\section{Generalities of 2-particle form factors of the stress-energy tensor}

Our primary interest is the zero particle form factor for the trace of the energy momentum tensor which determines $\rhovac$ according to 
\eqref{rhovacTrace}.    We wish to obtain $\rhovac$ from $2$ particle form factors.    In this section we  present some general, i.e. model independent,  properties of
these $2$-particle form factors in arbitrary spacetime dimension.   As stated in the Introduction,   this serves to fix the normalization of form factors in terms of physical masses.

Consider a relativistic QFT in $D=d+1$ spacetime dimensions.     
For simplicity  let us assume the theory consists of a single particle of mass $m$. 
One particle asymptotic scattering states are denoted as  $| \kvec \rangle$,  where $\kvec$ is the $d$-dimension momentum, and are on the mass shell with associated  
energy momentum $D$-vector $p_\kvec$: 
\beq
\label{4vector}
p_\kvec = (\omega_\kvec , \kvec ),  ~~~~~ \omega_\kvec = \sqrt{\kvec^2 + m^2}, ~~~~~ p_\kvec^2 = \omega_\kvec^2 - \kvec^2 = m^2.
\eeq
Since $d^d \kvec/\omega_\kvec$ is Lorentz invariant,  we define the resolution of the identity as follows: 
\beq
\label{UnityD}
{\bf  1 }  =  \sum_{n=0}^\infty \inv{n!} \int \frac{d^d \kvec_1}{ (2 \pi)^d \omega_{\kvec_1} } \cdots  \int \frac{d^d \kvec_n}{ (2 \pi)^d \omega_{\kvec_n}}
~
| \kvec_1, \kvec_2, \ldots \kvec_n \rangle \langle \kvec_1, \kvec_2, \ldots \kvec_n | 
\eeq
where 
\beq
\label{InnerProd}
\langle \kvec' | \kvec \rangle = (2 \pi)^d \, \omega_\kvec \, \delta^{(d)} ( \kvec' - \kvec ).
\eeq
The mass dimension\footnote{Henceforth all scaling dimensions refer to mass units.}   of the states $|\kvec\rangle$ is thus $(1-d)/2$.
For any quantum field operator $\CO (x) $  one can define its form factors 
\beq
\label{FFD}
F^\CO (\kvec_1, \kvec_2, \ldots, \kvec_n ) = \langle 0 | \CO |\kvec_1, \kvec_2, \ldots \kvec_n \rangle
\eeq
where in the above equation $\CO = \CO(0)$.   
Labeling particle states as $| p_\kvec \rangle$,  crossing symmetry reads 
\beq
\label{CrossingD} 
\langle p_{\kvec_1} | \CO (0) | p_{\kvec_2}, \ldots p_{\kvec_n } \rangle = \langle 0 | \CO(0) |-p_{\kvec_1}, p_{\kvec_2} , \ldots p_{\kvec_n} \rangle
\eeq

Let us now consider the operator $\CO = T_{\mu\nu}$,  the stress energy tensor.       
Since $\d^\mu T_{\mu \nu} =0$,  one can define an operator $A(x)$ where:
\beq
\label{TA}
T_{\mu\nu} (x) = \( \d_\mu \d_\nu - g_{\mu\nu} \d^2 \)\, A(x). 
\eeq
Define the trace of stress-energy tensor  $\TraceT \equiv  T_\mu^\mu$.
Then \eqref{TA} implies 
\beq
\label{TraceToo}
\langle 0|  \TraceT  |\kvec',  \kvec \rangle  = - d \, \, \frac{(p_{\kvec'}  + p_{\kvec} )^2}{(\kvec'+\kvec)^2 } ~  \langle 0 | T_{00}  | \kvec', \kvec \rangle 
\eeq
Using crossing symmetry: 
\beq
\label{TraceCrossed}
\langle \kvec' |  \TraceT   | \kvec \rangle  = - d \,\, \frac{(p_{\kvec'}  - p_{\kvec} )^2}{(\kvec'-\kvec)^2 } ~  \langle \kvec'  | T_{00}  | \kvec \rangle. 
\eeq
One can show that 
\beq
\label{crosslimit} 
\lim_{\kvec' \to \kvec}  
\frac{(p'-p)^2}{(\kvec'-\kvec)^2 } = -1.
\eeq
Thus 
\beq
\label{TraceToo2}
\langle \kvec |  \TraceT  | \kvec \rangle  = d \,  \langle \kvec| T_{00}  | \kvec \rangle .
\eeq
One also has 
\beq
\label{Intx} 
\int d^d \xvec \, \langle \kvec' | T_{00}  (x) |\kvec \rangle =   \int d^d \xvec \, e^{i(\kvec'-\kvec)\cdot \xvec} \, 
\langle \kvec' | T_{00} (0)  |\kvec \rangle = \langle \kvec' | H | \kvec \rangle = \omega_k  \, \langle \kvec' | \kvec \rangle
\eeq
where $H$ is the hamiltonian.   
This formally implies 
$ \langle \kvec | T_{00} (0) | \kvec \rangle = \omega_\kvec^2  $,      however based on \eqref{TraceToo2},   since the LHS is Lorentz invariant, 
$\omega_\kvec^2$ should be replaced by $m^2$.           One thus finds  simply 
\beq
\label{TraceFinal} 
\langle \kvec |  \TraceT  | \kvec \rangle =d\, m^2 ,
\eeq
which is independent of $\kvec$.  
  The above equation is what fixes the normalization of the $2$-particle form factors,  and this normalization is implicit for all the higher 
  $n\neq 2$ particle form factors in the bootstrap.  
The 1-particle form factors are also non-zero in general and by Lorentz invariance are constant:
\beq
\label{1particle}
\Lvac \TraceT | \kvec \rangle= {\rm constant} .
\eeq

It is important to note that  in obtaining \eqref{TraceFinal} we fixed by hand $\kvec' = \kvec$ under crossing symmetry in 
\eqref{TraceCrossed} based on Lorentz invariance.   This kind of manipulation of limits will be necessary below.    On the other hand the two-particle form factor $\Lvac \TraceT |\kvec, \kvec' \rangle$  contains a great deal more information on the underlying dynamics,   and the goal is to extract $\rhovac$ from it.      In the next section we propose a prescription for doing so in the case of integrable theories in $D=2$.

\section{$\rhovac$ from the 2-particle form factor for D=2 Integrable QFT}

Let us consider an integrable QFT in $D=2$ spacetime dimensions where the particles are labeled by ``a"  with physical mass $m_a$.   
We assume that each particle $a$ is its own anti-particle $\bar{a}$  for simplicity,   otherwise the formulas below involve charge conjugation matrices which
do not affect our main results.     Integrability implies no particle production so that the scattering of 2 particles leads only to 2-particle 
asymptotic states.    Also for simplicity,   we assume the S-matrix is diagonal,   which is to say that scattering of particles $a,b$ only produces particles of the same type,   thus one only need consider $S_{ab} = S_{ab}^{ab}$ matrix elements.

  \subsection{Review of form factor axioms}
  
In this subsection we give a brief review of the main formulas we will need.    For a more comprehensive review see for instance \cite{Smirnov,MussardoBook}.

The 2-particle form factors are functions of $s\equiv (p_1 + p_2)^2$ by Lorentz invariance.    For D=2, analytic properties of the S-matrix are more easily described in terms of the rapidity $\theta$:
\beq
\label{rapidities}
 p  = (\omega_{\kvec},  \kvec)  = (m\cosh \theta,  m \sinh \theta)
 \eeq
 which gives 
\beq
\label{s2D}
s = (p_a + p_b)^2 = m_a^2 + m_b^2 + 2 m_a m_b \cosh \theta_{ab}, ~~~~~ \theta_{ab} \equiv \theta_a - \theta_b . 
\eeq
One particle states are denoted as $|\theta\rangle_a$.    In the resolution of the identity \eqref{UnityD} one  has $\int d\kvec/\omega_\kvec = \int d\theta$,  thus 
  the resolution of the identity is now
\beq
\label{UnityD}
{\bf  1 }  =  \sum_{n=0; \{ a \} }^\infty \, \inv{n!} \int \frac{d \theta_1} {2 \pi }  \cdots  \int \frac{d \theta_n}{2 \pi} 
~
| \theta_1, \theta_2, \ldots \theta_n \rangle_{\{a_1 a_2 \ldots\}}  \,_{\{a_1, a_2 \ldots\}} \langle \theta_1, \theta _2, \ldots \theta_n | ,      ~~~~ _a\langle \theta' | \theta \rangle_b  = \delta_{ab} 2 \pi \delta(\theta-\theta') .
\eeq

For the remainder of this section $F$ refers to the form factors of $\TraceT$:
\beq
\label{FF2D1}
F_{a_1 a_2 a_3 \ldots} (\theta_1, \theta_2,  \theta_3, \ldots ) = \langle 0 | \TraceT  | \theta_1, \theta_2,  \theta_3, \ldots \rangle_{a_1 a_2 a_3 \ldots} 
\eeq
Since the states $|\theta\rangle$ have zero mass dimension,    all multi-particle form factors $F$ have the same mass dimension of $2$.
Form factors for any operator $\CO$ satisfy well-known axioms \cite{Smirnov}.    Although we are only interested in the 2-particle form factors,  for reasons that will be clear 
let us review these axioms for up to $4$ particles.   

\bigskip\bigskip

\noindent
{\it Crossing symmetry.}  ~~ $p \to - p$ corresponds to $\theta \to \theta + i \pi$ in \eqref{rapidities}.    This implies 
\beq
\label{FF2D3}
F_{a_1 a_2 a_3 a_4} ( \theta_1+ i \pi, \theta_2,  \theta_3, \theta_4 ) =
~ _{a_1}\langle \theta_1  | \TraceT  | \theta_2,  \theta_3, \theta_4 \rangle _{a_2 a_3 a_4} .
 \eeq

\bigskip\bigskip
\noindent
{\it  The Watson equation.} 

\beq
\label{FF2D2}
F_{a_1 a_2 a_3 a_4}  ( \theta_1, \theta_2,  \theta_3, \theta_4 ) =  S_{a_1 a_2}  (\theta_{12} )\,  F_{a_2 a_1 a_3 a_4} (\theta_2, \theta_1 , \theta_3, \theta_4 ).
\eeq

\bigskip\bigskip
\noindent
{\it  Kinematic poles.}    There are generic poles from the annihilation of a particle with its anti-particle which relates the $n$ particle form factor to that for $n-2$ particles:

\beq
\label{FF2D4}
-i \lim_{\theta_1 \to \theta_2 } (\theta_1 - \theta_2 ) ~  F_{a \bar{a}  a_3 a_4} ( \theta_1+ i \pi, \theta_2,  \theta_3, \theta_4 ) =
 \( 1 - S_{a a_3} (\theta_{23} ) \, S_{a  a_4 } (\theta_{24} )\) \, F_{a_3 a_4} (\theta_3, \theta_4) .
 \eeq

 \bigskip\bigskip
\noindent
{\it  Bound state  poles.}   If particle $c$ appears at a pole in $S_{ab} (\theta)$ at $\theta = i u_{ab}^c$ then there is a bound state with 
mass 
\beq
\label{bootmass}
m_c^2 = m_a^2 + m_b^2 + 2 m_a m_b \cos u_{ab}^c .
\eeq
The scattering of particle $c$ with any other particle $d$ satisfies the bootstrap equation
\beq
\label{bootS}
S_{cd} (\theta) = S_{ad} (\theta + i \bar{u}_{ac}^b ) \, S_{bd} (\theta - i \bar{u}_{bc}^a ),
\eeq
and the form factor also has such poles.   Starting with the S-matrix for the lightest particle $m_1$ with itself, the bootstrap can be closed, i.e. all other
 $m_a$'s and $S_{ab}$ can be determined.
 \bigskip\bigskip
 
 Since the above form factor axioms are linear in $F$,     they are incomplete since they do not fix the normalization of the $F$'s for the operator
 $\TraceT$.    Thus additional physical input is required.  
 In particular 
 \beq
 \label{constants}  
 \Lvac \TraceT \Rvac =  \rhovac/D,  ~~ {\rm and} ~~ \Lvac \TraceT |\theta\rangle_a \equiv F_a, ~~~~ {\rm are ~ constant}
 \eeq
 due to Lorentz invariance.   
   Such input must come from the UV since the two point correlation function of the 
 stress-energy tensor $\langle \TraceT (x) \TraceT (0) \rangle$  can be expressed in terms of form factors using the resolution of the identity 
 and if the theory is UV complete this correlation function is fixed at high energies by  the UV CFT.   
 We henceforth assume the QFT is UV complete.

 \subsection{A prescription for $\rhovac$}

 Suppose we are given the 2-particle form factors $F_{ab} (\theta_1, \theta_2 ) = F_{ab} (\theta_{12})$.     The kinematic pole equation \eqref{FF2D4} in general relates
 $n$-particle form factors the the $n-2$ ones.     However this equation makes no sense for $n=2$ as written.      However some version of it should be valid with some prescription,  and this is what we will present.      
 
 Consider the formal crossing relation
 \beq
 \label{prescrip1}
 ~_a \langle \theta_1 |  \TraceT | \theta_2 \rangle_b =  F_{ab} (\theta_1 + i \pi, \theta_2 ).
 \eeq
 The form factors are assumed to be normalized such that \eqref{TraceFinal} is satisfied,   which requires setting $\theta_1 = \theta_2 = \theta$ by hand and then using Lorentz invariance to conclude the result should be independent of $\theta$:
  \beq
 \label{prescrip2}
  ~_a \langle \theta |  \TraceT | \theta \rangle_a =  \lim_{\theta_1 \to  \theta_2} \, F_{aa} (\theta_1 + i \pi, \theta_2 ) = m_a^2 .
 \eeq
 This  properly normalizes the form factor with no additonal freedom to change it.   
 This kind of treatment of the rapidities will play an essential role for $\rhovac$ and the one point form factors $F_a$ below. 
 An important point is that if one does not  directly set $\theta_1 = \theta_2$ in \eqref{prescrip1},   there is still a lot of dynamical information in the form factor.      The second point is that one does not expect any kinematic poles as resonances for the vacuum $\Rvac$,  precisely because it is the vacuum.    We will show this explicitly below for the case of  the sinh-Gordon model.     Thus one does not need to cancel the pole in 
 \eqref{FF2D4}.       This leads us to propose the following.   The order of limits is important.     We first consider the high energy limit $\theta_1 \to \infty$ first,   then subsequently set  $\theta_1 = \theta_2$. We denote this order of limits as $\lim_{\theta_1 \to \infty|\theta_1 = \theta_2}$.
 We now  propose the prescription:
 \beq
 \label{prescrip3}
 - i \lim_{\theta_1 \to \infty|\theta_1 = \theta_2} \, F_{a\bar{a}} (\theta_1 + i \pi, \theta_2 ) =       
 \lim_{\theta_1 \to \infty|\theta_1 = \theta_2}   \( 1- S_{a\bar{a}} (\theta_{12} ) \)  \, \Lvac \TraceT \Rvac .
 \eeq
The above equation is a dramatic example of the mingling between the UV and the infra-red,  since the LHS is essentially fixed by 
masses that are measured at low energy,  whereas the RHS involves the extremely high energy limit of the S-matrix.

 The above formula \eqref{prescrip3} leads to a simple formula for $\rhovac$.    
In general the basic building blocks of  the two-body S-matrices $S_{ab} (\theta)$ 
are factors of $f_\alpha (\theta)$: 
\beq
\label{Sf}
S_{ab}  (\theta) = \prod_{\alpha \in \CA_{ab} }  f_\alpha (\theta), ~~~~~
f_\alpha (\theta) \equiv  \frac{ \sinh \half \( \theta + i \pi \alpha \)}{\sinh \half \( \theta - i \pi \alpha \)} ,
\eeq
where  $\CA_{ab} $ is a finite set of   $\alpha$'s.  
One has
\beq
\label{prescrip4}
\lim_{\theta \to  \pm \infty} f_\alpha (\theta) = e^{\pm i \pi \alpha} \( 1 \pm 2 i \sin (\pi \alpha) \, e^{-|\theta|} + \CO(e^{-2 | \theta |}) \).
\eeq
Thus 
\beq
\label{prescrip5} 
\lim_{\theta \to \infty} S_{ab} (\theta) = 1 + i \gcal_{ab} e^{-\theta} +  \CO(e^{-2 \theta}) , ~~~~~ \gcal_{ab} = \sum_{\alpha \in \CA_{ab}} \,
2 \sin \pi \alpha ,
\eeq
where we have used $\sum_{\alpha \in \CA_{ab}} \, \alpha = 0$.  
The prescription \eqref{prescrip3} combined with \eqref{prescrip2}  then leads to 
\beq
\label{prescrip6}
m_a^2 = \gcal_{aa} \,  \Lvac \TraceT \Rvac .
\eeq

 Since the vacuum has no quantum numbers,  the equation \eqref{prescrip6} must lead to the same result for 
 $\Lvac \TraceT \Rvac$  for any particle of type $a$.
 Namely any particle $a$ is able to probe $\Lvac \TraceT \Rvac$,  and this property was referred to as {\it particular democracy} in the 
 Introduction.        This is where the bootstap comes in.   
 Using that the  energy and momentum operators are  conserved quantities,   combined with the bootstrap equation
 \eqref{bootS},  one can show:\footnote{See for instance \cite{KlassenMelzer}.}
 \beq
 \label{bootg}
 \gcal_{ab} =  \mhat_a \mhat_b \,\,  \gcal_{11}  , ~~~~ \mhat_a \equiv m_a /m_1 ,
 \eeq
 where $\gcal_{11}$ is for the lightest particle $m_1$.   
Then \eqref{prescrip6} leads to,   for any particle $a$:
\beq
\label{TraceTfinal}
\Lvac \TraceT \Rvac = \frac{m_1^2}{\gcal_{11}}, ~~~~ \Longrightarrow ~~~ \rhovac = \frac{m_1^2}{2 \gcal_{11}}.
\eeq
The above result agrees with what is obtained from a thermodynamic approach,  namely the TBA \cite{ZamoTBA}.    This new derivation of 
$\rhovac$ is considerably simpler and doesn't depend on a finite temperature treatment.    As we will see,  this  has its advantages,   especially for its generalization to higher dimensions where a TBA treatment is not possible.

A related but slightly different  limit constrains the one-particle form factors.  Even for the stress-energy tensor,   these one-particle  form factors are constants and are not necessarily 
zero.    For instance for the magnetic perturbation of the Ising model,   $\TraceT$ is the spin field $\sigma$ and has non-zero
1-particle form factors \cite{DelfinoMussardo,DelfinoCardy,DelfinoE8}.     
 It was conjectured in these works  that when a subset of the rapidities are taken to infinity,
the form factors should factorize as a result of a kind of cluster decomposition.    Let us assume this conjecture.    In \eqref{prescrip3} if one only takes the $\theta_1 \to \infty$ limit without also the subsequent limit
$\theta_1 \to \theta_2$ the form factors should factorize into 1-particle form factors.       We thus propose:
\beq
\label{1point}
 \lim_{\theta_1 \to \infty} \, F_{ab} (\theta_1, \theta_2 ) =  \frac{F_a \, F_b}{\Lvac \TraceT \Rvac} ,
  ~~~~~ F_a \equiv  \Lvac \TraceT |\theta\rangle_a  ~ = {\rm constant}.
  \eeq
The above formula only makes sense if the LHS is finite and independent of rapidities.      The latter follows from 
\beq
\label{Slimit}
\lim_{\theta \to \infty }  S(\theta) = 1,
\eeq
which can be verified for the examples below.

 \subsection{Example:  Affine Toda theories}

The affine Toda theories provide a nice illustrative example since they have multiple particles of mass $m_a$ and the scattering is diagonal
(for real coupling $b$ below).  
They also have a coupling constant $b$ such that the exact result for $\rhovac$ can be compared with perturbation theory.   
We consider only the simply laced Lie algebras $G$,  the ADE series,   in particular $G=A_N = SU(N+1)$.   
Although the final result is known from the TBA,   we present it here as an illustration of the prescription \eqref{prescrip3} and 
{\it particular democracy}.          
 
\bigskip

Denote the simple roots  of  $SU(N+1)$ as $\alphavec_i$,  $i = 1, 2,  \dots N$:
\beq
\label{roots}
 \alphavec_i  =   (\alpha_i^1,  \alpha_i^2,  \ldots, \alpha_i^N ) \equiv    \{ \alpha_i^a, ~ a= 1,2, \ldots N \}   .
 \eeq
 Introduce $N$ real scalar fields $\phivec$:
 \beq
 \label{fields}
 \phivec = (\phi^1, \phi^2, \ldots, \phi^N) \equiv \{\phi^a, ~ a=1, 2, \ldots, N \}
 \eeq
 The untwisted affine Lie algebra  has one additional root 
 $ \alphavec_0 = - \sum_{i=1}^N \alphavec_i $. 
 With these definitions one can define the $2D$ Euclidean action  
 \beq
 \label{AffineAction}
 \CS = \int d^2 x \( \inv{8 \pi} \d \phivec \cdot \d \phivec  +  V(\phivec) \) , ~~~~~
 V (\phivec) = \mu \, \sum_{i=0}^N  e^{b \,  \alphavec_i \cdot \phivec } .
 \eeq
 The $1/8 \pi$ normalization corresponds to standard $2D$ conformal field theory conventions 
 where $\langle \phi^a (x) \phi^b (0) \rangle = - \delta_{ab} \log x^2 $ which fixes the convention for the coupling $b$. 
 With this convention $V(\phivec)$ is a strongly relevant perturbation of anomalous scaling dimension $-2 b^2$ if the all roots are conventionally normalized as   $\alpha_i^2 = 2$.    
 The masses satisfy the relation
 \beq
 \label{affineMasses}
\mhat_a  =  \frac{m_a}{m_1} = \frac{\sin (a \pi/(N+1))}{\sin (\pi/(N+1)}. 
\eeq
 
 The affine Toda theories have many applications.      For real coupling $b$ the S-matrices were first obtained in \cite{Fateev}.  For $SU(2)$ the model is the sinh-Gordon model.      The more interesting physical applications are for purely imaginary coupling $b$,  where for $SU(2)$ this is the sine-Gordon model.    For the latter the scattering is not diagonal except at the so-called reflectionless points.    For other Lie groups  $G$,  the affine Toda theories have solitons in the spectrum and the scattering is not diagonal and satisfies the Yang-Baxter equation \cite{BernardLeClair,Hollowood}.    The quantum theory for imaginary $b$ has a quantum affine symmetry 
 $\CU_q  (\hat{G})$ where $\hat{G}$ is the affine extension of $G$ \cite{BernardLeClair}.     For $q$ a root of unity,   some RSOS restrictions of these theories 
 describe integrable perturbations of minimal CFT's \cite{AhnBL}.     The  RSOS restricted $SU(3)$ theory describes the 3-state Potts model.    Restriction of the $G=E_8$ theory describes
 magnetic perturbations of the Ising model and one can reproduce the results of Zamolodchikov \cite{ZamoE8},   where  the theory is diagonal under RSOS restriction.       
 For the present article we only consider real coupling $b$ since the theory is diagonal for arbitrary $b$.    A comprehensive review of such theories can be found in \cite{Braden}.    
 
The S-matrix bootstrap can be completed starting with the just the lightest mass particle of mass $m_1$.  
Let $S_{11} (\theta)$ denote the S-matrix for this particle  with itself:
\beq
\label{S11}
S_{11} = \prod_{\alpha \in \CA_{11} } f_\alpha =  f_{\tfrac{2}{h}} \, f_{- \tfrac{2 \gamma}{h}} \,  f_{\tfrac{2(\gamma -1)}{h}},   ~~~ h = N+1, ~~~~~~ \gamma 
\equiv \frac{b^2}{1+b^2} .
\eeq 
($h$ is the dual Coxeter number.) 
Then
\beq
\label{Toda1}
\gcal_{11}  = 2 \sum_{\alpha \in \CA_{11}}  \sin \pi \alpha  = 
 - 8 \sin(\pi/h)  \sin (\pi \gamma/h ) \sin (\pi(1-\gamma)/h) . 
\eeq
Based on equation \eqref{TraceTfinal},   
one then finds
\beq
\label{gcalToda2}
\rhovac =  - \frac{m_1^2}{16 \, \sin (\pi/h ) \, \sin ( \pi \gamma/h ) \, \sin ( \pi (1-\gamma)/h )}.
\eeq
Expanding $\rhovac$ for $N=1$ (the sinh-Gordon model) one finds 
\beq
\label{shGasym}
\lim_{b\to 0} \, \rhovac = - m^2 \(  \inv{8 \pi b^2} - \inv{8 \pi} - \frac{\pi b^2}{48} + \frac{\pi b^4}{48} - \frac{(60 + 7 \pi^2) \pi b^6}{2880} + \ldots \)~~~~~(N=1). 
\eeq
Rescaling $m \to b^2 m$ the above defines a well-defined perturbative expansion that can be compared with Feynman diagram perturbation 
theory,   which was understood by Destri-deVega by summing over all tadpole Feynman diagrams in a rather complicated calculation \cite{DestriDeVega}.

 \subsection{Analytical details for the sinh-Gordon model}
 
The detailed analysis of this section is not essential for our purposes since we obtained \eqref{TraceTfinal} without knowing 
the exact 2-particle form factor function,   but it's useful to analyze an example where the 2-particle form factors are explicitly known 
so as to verify some of the above properties.  
The affine Toda theory for $SU(2)$ is the sinh-Gordon model with the action 
\beq
\label{sinhGaction}
\CS = \int d^2 x \( \inv{8 \pi} (\d_\mu \phi \, \d^\mu \phi ) + 2 \mu \cosh  ( \sqrt{2}\, b\,  \phi ) \) .
\eeq
The spectrum consists of a single particle of physical mass $m$.
From \eqref{S11} one has 
\beq
\label{sinhGSmatrix}
S (\theta) =  - f_{-\gamma} (\theta)  f_{\gamma -1} (\theta)  = \frac{ \sinh \theta - i \sin \pi \gamma}{\sinh \theta + i \sin \pi \gamma}, ~~~~~ \gamma \equiv b^2/(1+b^2) .
\eeq
This already  implies
\beq
\label{rhovacShG}
\gcal = -4 \sin \pi \gamma,  ~~~~~\Longrightarrow ~~\Lvac \TraceT \Rvac = - \frac{m^2}{4 \sin \pi \gamma} = 2 \rhovac .
\eeq
and $\rhovac$ is negative.

Form factors for the sinh-Gordon model were studied in great detail by Fring, Mussardo and Simonetti \cite{FringMussardo}.    
Henceforth $F(\theta_1, \theta_2) = F(\theta_{12})$ is the two particle form factor for $\TraceT$.       
From the Watson equation we define a minimal solution $\Fmin$:
\beq
\label{Fmineq}
\Fmin (\theta) = S(\theta) \, \Fmin (-\theta) .
\eeq
Form factors for $\CO$  for $n>2$ numbers of particles factorize into a product of $\Fmin (\theta_{ij} )$ times a polynomial in $e^{\theta_i}$,  where the polynomial 
depends on the operator $\CO$.   
$\Fmin$  has the integral  representation\footnote{For numerical evaluation of this integral,  the  hybrid formula  
eq. 4.36 in \cite{FringMussardo} is very useful.     See also \cite{Negro} for some more recent analysis of $\Fmin$.}:
\beq
\label{FminInt}
\log \Fmin (\theta) = -4 \int_0^\infty \frac{dx}{x} ~ \frac{ \sinh (x \gamma/2) \sinh(x(1-\gamma)/2 ) \sinh(x/2)}{\sinh^2 (x)} \, \cdot 
\cos (\thetatilde x/\pi ), ~~~~~~~~\thetatilde \equiv i \pi - \theta .
\eeq
Note that $\Fmin(\theta + i \pi)$ is real if $\theta$ is real.  
The properly normalized form factor is
\beq
\label{Trace1} 
F(\theta_{12}) = \langle 0 | \TraceT  | \theta_1, \theta_2 \rangle = m^2 \,  \frac{\Fmin (\theta_{12} )}{\Fmin (i \pi )}.
\eeq
With this normalization 
\beq
\label{Trace3}
\langle \theta  | \TraceT | \theta  \rangle =
 \lim_{\theta \to 0} \frac{\Fmin (\theta + i \pi )}{\Fmin (i \pi )} = m^2.
\eeq

Let us first show that there is no kinematic resonance poles in $F(\theta_1 + i \pi, \theta_2) = F(\theta_{12} + i \pi )$, as anticipated above,  such that
the residue axiom \eqref{FF2D4} makes no sense for the 2-particle form factor.  
For this purpose the following identity,   which is specific to the sinh-Gordon model,  is very useful:
\beq
\label{sinhGIdentity}
\Fmin (\theta + i \pi ) \, \Fmin (\theta) = \frac{\sinh \theta}{\sinh \theta +i \sin \pi \gamma} .
\eeq
Using this 
\beq
\label{res3}
F(\theta + i \pi)   = 
\inv{\Fmin (\theta) \Fmin(i \pi)}  \, \frac{\sinh \theta}{\sinh \theta +i \sin \pi \gamma} .
\eeq
From $S(0) = -1$ one has $\lim_{\theta \to 0} \Fmin (\theta) = 0$.   
In order to obtain a series in powers of $\theta$ we consider 
\beq
\label{Fdiv}
\d_\theta \log \( \frac{ \Fmin (\theta)}{\Fmin (-\theta)} \) = \d_\theta \log S(\theta) .
\eeq
From this one can show 
\beq
\label{Fminzero}
\lim_{\theta \to 0} \Fmin (\theta) = \inv{\Fmin(i \pi) \sin (\pi \gamma)} \( -i \theta + \frac{\theta^2}{\sin(\pi \gamma) } + \ldots \).
\eeq
Thus as expected 
\beq
\label{res4}  
 \lim_{\theta \to 0}  F(\theta + i \pi)  = m^2.
\eeq
One can also show for large $|\theta|$: 
\beq
\label{LargeTheta}
\lim_{\theta \to \pm \infty}  \Fmin(\theta) \approx 1+   \frac{i \gcal }{4\sinh \theta} .
\eeq

It is also interesting to to study the coupling constant dependence of the basic constant $\Fmin (i \pi)$.
From the integral representation \eqref{FminInt} one can show 
\beq
\label{FminIpi}
\log \Fmin ( i \pi ) = -\gamma + \frac{\gamma^2 \pi^2 }{8} - \frac{ \gamma^3 \pi^2 }{18} +  \frac{\gamma^4 \pi^4}{192} + \CO (\gamma^5 ).
\eeq
At self-dual point $b=1$: 
\beq
\label{FminSelfDual}
\log \Fmin (i \pi)_{\gamma = 1/2}  = - \frac{2\, \CG}{\pi} + {\rm arccoth} (3) , ~~~~ \Longrightarrow ~~ \Fmin(i \pi)_{\gamma = 1/2}  =  0.789348.. ...
\eeq
where $\CG = 0.915966$ is the Catalan constant\footnote{$\CG = i ( \Li_2 (-i ) -  \Li_2 (i) )/2$,   where $\Li_2$ is a poly-logarithm.}.

Finally let us mention that $\rhovac$ can also be determined from a $4$-particle form factor.   Define the kernel $G(\theta)$ which 
appears in the TBA:
\beq
\label{Gkernel}
G(\theta) = -i \d_\theta \log S (\theta) = \frac{2 \cosh \theta \sin \pi \gamma}{\cosh^2 \theta - \cos^2 \pi \gamma }.
\eeq
Then one can show \cite{LeClairMussardo} 
\beq
\label{4particle}
\langle \theta_2, \theta_1 | \TraceT | \theta_1, \theta_2 \rangle = 2 m^2 \, G(\theta_{12} ) \, \cosh (\theta_{12} ).
\eeq
Thus one has 
\beq
\label{4particle2}
\lim_{\theta_1 \to \theta_2 } \, \langle \theta_2, \theta_1 | \TraceT | \theta_1, \theta_2 \rangle = \frac{4 m^2}{\sin (\pi \gamma)} =  -32 \, \rhovac .
\eeq

\subsection{Summary}

Before turning to higher dimensions D in the next section,   it's useful to summarize the main points of this section.  
The main new result is the formula \eqref{prescrip3} which is a prescription for obtaining $\rhovac$ from the 2-particle form factor using the 
bootstrap,  and involves a delicate limit where the order of limits matters.  
Due to the required properties of the crossed form factor,  the LHS of this equation is simply $m^2$ and this leads to the formula \eqref{TraceTfinal}.
The check of this prescription is that it reproduces the previously known formula from the TBA.   Although the formula \eqref{TraceTfinal} was derived 
for diagonal theories,   it also applies to non-diagonal scattering,  as illustrated for the sine-Gordon model in \cite{Mingling}.    
In obtaining \eqref{TraceTfinal} it's important to note that the detailed formula for the 2-particle form factor was not necessary and the final result only depends on S-matrix parameters.     We also proposed the principle of {\it particular democracy} and showed how it follows from consistency of the S-matrix bootstrap.     For the sinh-Gordon model we provided additional analytical details supporting the prescription \eqref{prescrip3}.     
These results imply that UV divergences in $\rhovac$ for free theories can be cured with interactions,    where the free field theory limit is 
$\gcal \to 0$.

%%%%%%%%%%%%%%%%%%%%%%%%%%%%%%%%%%%%%%

\section{$\rhovac$ from the form factor bootstrap in higher dimensions}

In this section we will propose a prescription for determining $\rhovac$ from the 2-particle form factors of  $\TraceT$ in D=4 spacetime dimensions.   
The  D=2 result in the last section relied on integrability,   which is not available  in higher dimensions.    In particular for the D=2 theories integrability implies that there is no particle production,    the S-matrix factorizes,   and this leads to a kind of factorization for $n>2$ particle form factors. 
Integrability also entails the TBA where our prescription  \eqref{prescrip3} for $\rhovac$ can be checked by comparison with the Thermal definition of 
$\rhovac$ proposed in \cite{ALCC}.     
None of these integrability features exist for D=4.   However bootstrap ideas for massive theories in principle apply to higher dimensions, 
though in a much more complicated fashion \cite{Paulos1,Homrich,Karateev}.    Moreover,   since our prescription is based only on the 2-particle form factor,   one does not necessarily need factorization of the S-matrix nor the factorization of the  form factor.      
This opens the possibility that $\rhovac$ can be determined from a form factor bootstrap equation based on the 2-particle form factor, 
and this is the primary focus of this section.    In the next subsection we review the basic constraints on the 2-particle form factor for $\TraceT$,  which only depends on the Mandelstam variable $s$.      We then re-write the D=2 result \eqref{prescrip3} in terms of $s$.     
The latter  result has a natural generalization for $\rhovac$ in $D=4$ which we present below.

\subsection{Generalities}

For simplicity we assume the theory consists of a single particle of mass $m$ which is its own anti-particle.    
Throughout this section $F(\kvec_1 , \kvec_2 )$ refers to the 2-particle form factor for the trace of the stress energy tensor $\TraceT$ and 
and $F_1$ the 1-particle form factor which is  constant:    
\beq
\label{Fdef}
F(\kvec_1, \kvec_2 ) \equiv \Lvac \TraceT | \kvec_1 , \kvec_2 \rangle,   ~~~~ F_1 = \Lvac \TraceT | \kvec \rangle = {\rm constant} .
\eeq
Since $\TraceT$ is Lorentz invariant $F(\kvec_1, \kvec_2) = F(s)$ where $s = (p_1 + p_2)^2$:
\beq
\label{sD}
s = (p_1 + p_2)^2 = 2m^2 + 2 \( \omega_{\kvec_1} \omega_{\kvec_2} -   k_1 k_2 \, \cos \phi \), ~~~~ \hat{\kvec}_1 \cdot \hat{\kvec}_2  = \cos \phi . 
\eeq
where $\phi$ is the angle between $\kvec_1$ and $\kvec_2$.    In the center of mass frame,  
$\cos \phi = -1$ and   $|\kvec_1| = |\kvec_2| = |\kvec|$,  thus 
$s=4 m^2 + 2 \kvec^2$.    

In any dimension $D$ one still has the Watson equation:
\beq
\label{WatsonD} 
F(s) = S(s) \, F^* (s) +  \ldots,
\eeq
where $S(s)$  is the S-matrix  for the angular momentum $\ell = 0$ partial wave.     The additional terms $\ldots$ refer to contributions from thresholds for the production of more than 2 particles,  starting at $s>(3 m)^2$.     These are absent for integrable theories in  D=2  since there is no particle production,  and \eqref{WatsonD}  is equivalent to \eqref{FF2D2} since $F^* (\theta) = F(-\theta)$ and $S^* (\theta) S(\theta) = S(-\theta) S(\theta) =1$ due to unitarity.     For the remainder of this work we neglect the  effect of higher  terms in \eqref{WatsonD}.    One justification for this is 
that in $2D$  our prescription for $\rhovac$ did not depend on detailed knowledge of the 2-particle form factor.  
The optical theorem can relate $\Im \, F(s)$ to the 1 particle form factors, and this implies generic poles at $s=m^2$.    See for instance  the arguments in \cite{Karateev}.   However we are interested in bootstrapping 
$F(s)$ {\it down} to the zero-particle form factor $\Lvac \TraceT \Rvac$,  skipping over the intermediate 1-particle form factor,  although we will return to it below\footnote{The work \cite{Karateev} focussed mainly on 2 and 1-particle form factors but not the 0-particle form factor considered here.}.

\subsection{D=2  case in terms of Mandelstam variable $s$}

In this sub-section we express our prescription \eqref{prescrip3} in terms of the Mandelstam variable $s$:
\beq
\label{s2D}
F(s) \equiv \Lvac \TraceT | \kvec_1, \kvec_2 \rangle, ~~~~~~ 
s= (p_1 + p_2)^2 =  2 m^2 \( 1 + \cosh \theta_{12} \).
\eeq
Crossing symmetry $p_1 \to - p_1$ implies 
\beq
\label{crossings}
s \to 4 m^2 - s. 
\eeq
Thus \eqref{TraceTfinal} implies 
\beq
\label{CrossedF}
\langle \kvec | \TraceT | \kvec \rangle =  \lim_{s \to 4 m^2}  F(4m^2 -s ) = m^2.
\eeq

The high energy limit of the S-matrix in terms of $s$ is 
\beq
\label{HighS}
\lim_{\theta \to \infty}  S(\theta) =  1 + i\frac{ \gcal}{2 \sinh \theta}  = 1+ i \frac{\gcal \, m^2}{\sqrt{s (s-4m^2)}}.
\eeq
As expected,  it has square-root branch cuts originating at $s=0$ and $4 m^2$.   
This gives 
\beq
\label{HighS2}
\lim_{s \to \infty}  (1- S(s))   = -i\,  \frac{\gcal \, m^2}{s} .
\eeq
Now one has 
\beq
\label{slim}
\lim_{\theta_1 \to \infty |\theta_1 = \theta_2} \, s =  m^2 .
\eeq
Thus the prescription \eqref{prescrip3} reads 
\beq
 \label{BootD}
 -i \, m^2 = 
 - i \lim_{s\to \infty|s=4 m^2} \, F(4m^2 -s)  =       
  \lim_{s\to \infty|s= m^2}    \( 1- S (s) \)  \, \Lvac \TraceT \Rvac .
 \eeq
This leads to the result in Section III:
\beq
\label{rhovacs}
\Lvac \TraceT \Rvac =  \frac{m^2}{\gcal}.
\eeq

\subsection{Prescription for $\rhovac$ in D=4 spacetime dimensions} 

Here one has to work with standard kinematic variables rather than the rapidity,   in particular $s$.
In the complex $s$-plane,   it is known that there are two square-root branch cuts along the real axis,  one along the negative axis $s<0$ and another 
above  the 2-particle threshold $s>4 m^2$.    
Written in terms of $s$,   the equation \eqref{BootD} has a natural generalization to higher dimensions which we propose below. 
In D=4 dimensions,   the states $|\kvec \rangle$ have  dimension $-1$.    Since the operator $\TraceT$ has dimension $4$,  
$\Lvac \TraceT \Rvac$ has dimension $4$,   the two-particle form factor $F(s)$ has dimension $2$ and the 1-particle form factors 
$F_1$ have dimension $3$.

\def\That{\tilde{\CT}}

As operators 
\beq
\label{Tdef}
S = 1 + i \CT
\eeq
where $\CT$ contains the interactions.    
The D=2 result \eqref{prescrip3}  involves $1-S$,  thus we consider the {\it diagonal} $\CT$ operator matrix element:
\beq
\label{That}
\CT (s)  \equiv   \langle \kvec_1, \kvec_2 | \CT |\kvec_1 , \kvec_2 \rangle .
\eeq
Since $\CT$ is dimensionless,    $\CT (s)$ has dimension $-4$.   In the center of mass frame,  the above matrix element is independent of 
the angle $\phi$ in \eqref{sD} since the momentum in the $\langle {\rm  bra} |$  and $| {\rm ket}\rangle$  are the same.    Thus for the matrix element $\CT (s)$ one can consider both the incoming and outgoing particles along the same line in the center of mass frame.       
The fact that the $\langle {\rm bra}| $ and $|{\rm ket}\rangle$ states in \eqref{That}  are the same is to be expected:   In the Thermal approach, 
${\rm Tr} e^{- \beta \, H}$ involves integrating over matrix elements such as in \eqref{That}.   
In D=2 dimensions,   one-point correlation functions of fields  at finite temperature involves such matrix elements \cite{LeClairMussardo}. 
These diagonal matrix elements are also central to quantum statistical mechanics formulated completely in terms of the S-matrix \cite{Dashen}.
\beq
\label{dashen} 
Z = Z_0 + \inv{2\pi} \int_0^\infty  dE\, \, e^{-\beta E}\, {\rm Tr} \,   \Im \( \d_E \log \hat{S} (E) \).
\eeq

Returning to our problem,   based on the correct D=2 result \eqref{BootD} we propose 
\beq
 \label{Boot4D}
  \lim_{s\to \infty|s=4 m^2} \, F(4m^2 -s)  =       
 m^2 \(   \lim_{s\to \infty|s= m^2}   \CT(s)  \) \, \Lvac \TraceT \Rvac .
 \eeq
 The extra factor of $m^2$ on the RHS will serve to define a dimensionless coupling constant $\gcal$ below. 
 Now the LHS of the above equation  equals $3 m^2 $ by \eqref{TraceTfinal}.     
 Since $\CT (s)$ has dimension $-4$,  let us assume 
\beq
\label{Tlarges}
\lim_{s\to \infty }  \CT(s) = \frac{1}{m^2} \, \frac{\gcal}{s} .
\eeq
Then \eqref{Boot4D} implies 
\beq
\label{rhovac4D}
\Lvac \TraceT \Rvac =  \, \frac{3 m^4}{\gcal}  ~~~~ \Longrightarrow ~~~ \rhovac = \frac{3}{4}   \, \frac{m^4}{\gcal}  .
\eeq
As for the D=2 case,  the explicit function $F(s)$ was not needed in order to obtain the above result.    

Let us turn now to the 1-point functions,  and as before label the particle type as ``a".     
Then the formula \eqref{1point},  which was based on a cluster decomposition and didn't rely on integrability,  is still expected to hold:
\beq
\label{1point4D}
 \lim_{s\to \infty } \, F_{ab} (s) =  \frac{F_a \, F_b}{\Lvac \TraceT \Rvac} ,
  ~~~~~ F_a \equiv  \Lvac \TraceT |\theta\rangle_a  ~ = {\rm constant}.
  \eeq
 The above equation is dimensionally correct,   however it can only be valid if $\lim_{s \to \infty}  F_{ab} (s)$ is finite.    In $2D$ this is ultimately a consequence of \eqref{Slimit}.

Interestingly,   the result \eqref{rhovac4D}  is close in spirit to a bound on $\rhovac$ which was obtained by  very different considerations involving 
 Swampland ideas in connection with electrically charged black holes \cite{Montero1,Montero2}.   There it was proposed that 
\beq
\label{Montero}
\rhovac <    \frac{m^4}{2 e^2}
\eeq
where $m$ is the mass of a charged particle,  and $\alpha = e^2/4 \pi  $  is the electromagnetic  fine structure constant.  
Whereas the power $m^4$ is expected just based on dimensional analysis,      
the interesting observation is that it is also inversely proportional to an interaction coupling $\gcal$.   Let us add that  for our proposal for $\rhovac$ we did not need to assume the particle is electrically charged.      

In closing this section,   we wish to emphasize that we have not rigorously derived our prescription \eqref{Boot4D},  nor have we have completely justified the high energy limit \eqref{Tlarges}.      Rather we based \eqref{Boot4D} on generalizing the D=2 result \eqref{prescrip3} in the most natural manner under certain assumptions that we stated.      Nevertheless  \eqref{Boot4D} should be viewed as conjectural.

\section{Potential applications to  Beyond the Standard Model physics}

Considering  the results of the last section,  we ask the reader to  allow us to take the liberty to speculate on potential implications for physics beyond the Standard Model of particle physics,   despite our lack of expertise in its voluminous detailed  intricacies.   
A major open question is the origin of the cosmological constant $\rho_\Lambda$,   i.e. Dark Energy,  which has been measured to be positive and unexplainably small \cite{Weinberg,Martin,Carroll,WMAP}.   There is not yet a consensus on the origins of $\rho_\Lambda$ nor its likely resolution.     However one prominent idea \cite{Weinberg},  which is at the origin of the cosmological constant problem,  is that it arises from 
the zero point energy of quantized fields,  namely $\rhovac$ considered in this paper.\footnote{For our perspective on the Cosmological Constant Problem,  which is not original,    see the short discussion in \cite{ALCC} and references therein.}     There actually has not been much serious effort toward  a proper calculation of $\rhovac$ in the past literature,  and one still encounters the  incorrect statement  that the Standard Model predicts  that  it is proportional to 
a cut-off equal to the Planck scale to the $4$-th power.\footnote{The latter is the origin of the  still often quoted statement that the prediction is off
by $120$ orders of magnitude,  which is incorrect even in theory.}     Let us suppose then that $\rhovac$   is the only source for 
$\rho_\Lambda$.         Then our understanding of $\rhovac$ gained in this article has some potential implications for physics beyond the Standard 
Model.

\def\mHiggs{M_{\rm Higgs}} 

The Standard Model of particle physics is based on the $SU(3) \otimes SU(2) \otimes U(1)$ Yang-Mills theory,    
where the  $SU(3)$ is QCD.      
At the electroweak scale,  on the order of $ \mHiggs \approx 125\, {\rm GeV}$,  the $SU(2) \otimes U(1)$ is spontaneously broken  (SSB) by the Higgs mechanism 
and this is the standard proposed origin of {\it all} particle masses,  both for the  the quarks and leptons.   
Simplifying matters,   neutrinos only have a  left-handed handed helicity, 
which protects them from obtaining a mass from the Higgs mechanism.

\bigskip

To clarify our reasoning,   let us itemize our assumptions,  since  they certainly require more scrutiny. 
    
\bigskip
\begin{assumption}

We assume that  the Standard Model is UV complete and in the deep UV it is a CFT.    It helps to assume  the UV theory is asymptotically free 
like  QCD,  which would imply   $\lim_{s  \to \infty}  S(s) = 1$,  however it's possible this could be relaxed.

\end{assumption} 

\bigskip

\begin{assumption}

 Since $\rhovac$ is fixed by the deep UV,   we assume any Spontaneous Symmetry Breaking (SSB)   far below the  higher  energy   scale  that determines 
 $\rhovac$,   
such as electroweak symmetry breaking or the de-confinement transition in QCD,   does not play a role in determining $\rhovac$ since the latter is determined at potentially much higher energies.       On the contrary,  if such SSB scales were  important to $\rhovac$,  then it is already known that these energy  scales are much too high to explain the measured cosmological constant $\rho_\Lambda$.   
To justify this,  let us mention that we provided an estimate of $\rhovac$ for QCD with 3 massive quarks  based on lattice calculations without assuming the de-confinement transition.     

\end{assumption}

\bigskip

\begin{assumption}

We assume that in the {\it very}  deep UV,   way above the SSB Higgs scale $\mHiggs$,   all particles which are thought to obtain their  mass from the Higgs mechanism are massless.    We mention that our calculation of $\rhovac$ for QCD already assumed non-zero quark masses from the Higgs mechanism,   and this gave a value of $\rhovac$ this is finite and well-defined,  but much too high to explain $\rho_\Lambda$.      

\end{assumption}

\bigskip

\begin{assumption}

 In the deep UV,    since $\Lvac \TraceT \Rvac =0$ for a CFT,   a non-zero $\rhovac$ must arise from at least one  non-zero mass particle 
of mass $\mgod$ which cannot be explained by the Higgs mechanism.  It must then arise from a relevant perturbation of the UV CFT which sets a mass scale.     This perturbed  theory should not be supersymmetric otherwise $\rhovac =0$ by the usual arguments.    This does not rule out that the UV CFT is supersymmetric,  since the relevant perturbation could break it explicitly.       Then the above results of this paper would lead us to propose 
\beq
\label{rhovacgod}
\rhovac  \approx \frac{3}{4}   \frac{\mgod^4}{\gcal} ,
\eeq
where $\gcal$ is a dimensionless interaction coupling.   
For lack of a better name,  let us refer to this hypothetical particle as the  zero-on,  where ``zero"  refers to the vacuum,   or simply the  $\zeron$. 
This particle is presumed stable,   otherwise there would be imaginary parts to poles in $s$ signifying a decay rate which we have not considered.

\end{assumption}

\bigskip

The measured value of the cosmological constant is \footnote{See for instance \cite{WMAP}.}
\beq
\label{rhovacNeutrino}
\rho_\Lambda \approx  10^{-9} {\rm Joule}/{\rm meter}^3 \approx (0.003 \,{\rm eV})^4 . 
\eeq
Interestingly,   there already exists very good candidate for the $\zeron$,  which is  the neutrinos.      In the Standard Model they only occur with
left-handed helicity,  thus it is thought that a non-zero neutrino mass cannot arise from the Higgs mechanism.   On the other hand,   if the neutrino is its own anti-particle, namely a Majorana fermion,   since the anti-particle is right-handed,  it can pair with the left-handed ones to accommodate a mass term in the action.   
If this is the case,   one expects $\gcal$ to be on the order of the fine structure constant $e^2/\hbar c \approx 1/137$ for the electro-weak sector.  
Neglecting overall constants of order $1$,  one sees from \eqref{rhovacNeutrino} that if $\mgod \sim 0.001\, {\rm eV}$ then this could account for the cosmological constant,  and $\mgod$ is close to proposed Majorana neutrino masses \cite{Nir}.

If the zeron is indeed a massive Majorana fermion,    then the  formula \eqref{rhovac4D} 
implies that the Cosmological Constant can be measured from co-linear  neutrino-anti-neutrino scattering by measuring $\gcal$ based on 
\eqref{Tlarges},     but only at energies much higher than the electro-weak SSB scale $\mHiggs$.    The principle of {\it particular democracy} implies $\rhovac$ can be measured from the scattering of any of such neutrinos.\footnote{Incidentally,  
   massive Majorana neutrinos  are a leading candidate to explain the matter/anti-matter asymmetry in the current Universe 
since it can lead to  CP violation \cite{CPMajorana}.
Roughly speaking,   a massive Dirac fermion can be viewed as two CP conjugate Majorana pairs.    It would be quite remarkable if Majorana neutrinos 
could  both explain CP-violation and the Cosmological Constant.    Clearly more scrutiny of this idea is worthy of consideration.}

\section{Conclusions}

A main results of this article are  the  prescription  \eqref{prescrip3} for completing the form factor bootstrap for the trace of the stress-energy tensor,
  in particular how to obtain 
the zero particle form factor which determines $\rhovac$  from the 2-particle one,  and the proposal \eqref{Boot4D} for its generalization to D=4.    
This completion requires additional physical input at very high energies which can be determined from the S-matrix.    For integrable QFT's in 
D=2 spacetime dimensions,  our prescription reproduces previously known exact results from the thermodynamic Bethe ansatz.    
   In these even dimensions one finds the simple expression 
$\rhovac \propto m^\Dsmall/\gcal$ where $m$ is a physical mass and $\gcal$ a dimensionless  interaction coup ing which can be inferred from the high energy behavior of the S-matrix.    The density $\rhovac$ diverges as the interaction $\gcal \to 0$ due to known and well-understood UV divergences in free theories.
This  implies that interactions can potentially cure the known  UV divergences in free QFT's.

With the understanding we obtained from these considerations,   we speculated on potential applications to the Standard Model of particle physics. 
Assuming that the cosmological constant $\rho_\Lambda$ comes solely from the vacuum energy density studied  in this article,   then  this would seem to imply the existence of a massive particle that does not obtain its mass from SSB of the electro-weak sector,   which we termed the zeron. 
Massive Majorana neutrinos are  a strong candidate for the zeron,  since previously proposed  neutrino masses  approximately have  the correct magnitude to account for the astronomically observed $\rho_\Lambda$ if $\gcal$ is of order 1.

%%%%%%%%%%%%%%%%%%%%%%%%%%%%%%%%%%%%%%%%%%%%%%%%%%%%%%%%%%%%%%%%%%%%%%%

\section{Acknowledgements} 

We wish to  thank Peter Lepage,  Giuseppe Mussardo and Matthias Neubert for discussions.

\end{document}